\documentclass[12pt,preprint]{aastex}

\newcommand{\teff}{$T_{\mathrm{eff}}$}
\newcommand{\hii}{H {\small II}}
\newcommand{\fei}{[\ion{Fe}{1}/H]}
\newcommand{\feii}{[\ion{Fe}{2}/H]}
\newcommand{\dfe}{$\Delta \mathrm{Fe}$}
\newcommand{\otrip}{[O/H]$_{\mathrm{Trip}}$}
\newcommand{\cairt}{$\log \mathrm{R}^{\prime}_{\mathrm{IRT}}$}
\newcommand{\halpha}{$\log \mathrm{R}^{\prime}_{\mathrm{H}\alpha}$}

\begin{document}

\title{\ion{Fe}{1} and \ion{Fe}{2} ABUNDANCES OF SOLAR-TYPE DWARFS IN THE 
       PLEIADES OPEN CLUSTER\altaffilmark{1}}

\author{Simon C. Schuler\altaffilmark{2,3}}
\affil{National Optical Astronomy Observatory/Cerro Tololo Inter-American Observatory}
\affil{Casilla 603, La Serena, Chile}
\email{sschuler@noao.edu}

\author{Adele L. Plunkett\altaffilmark{4,5}}
\affil{Department of Physics, Middlebury College}
\affil{McCardell Bicentennial Hall, Middlebury, VT 05753}
\email{aplunket@middlebury.edu}

\author{Jeremy R. King}
\affil{Department of Physics and Astronomy, Clemson University}
\affil{118 Kinard Laboratory, Clemson, SC 29634}
\email{jking2@ces.clemson.edu}

\and

\author{Marc H. Pinsonneault}
\affil{Department of Astronomy, Ohio State University}
\affil{140 West 18th Avenue, Columbus, OH  43210}
\email{pinsono@astronomy.ohio-state.edu}

\altaffiltext{1}{Based on observations obtained with the High Resolution Spectrograph on 
                 the Hobby - Eberly Telescope, which is operated by McDonald Observatory 
		 on behalf of the University of Texas at Austin, the Pennsylvania State 
		 University, Stanford University, the Ludwig-Maximillians-Universitaet, 
		 Munich, and the George-August-Universitaet, Goettingen. Public Access 
		 time on the Hobby - Eberly Telescope administered through the National
		 Optical Astronomy Observatory was made possible through an agreement 
		 with the National Science Foundation.}
\altaffiltext{2}{Current address: NOAO, 950 North Cherry Avenue, Tucson, AZ  85719}
\altaffiltext{3}{Leo Goldberg Fellow}
\altaffiltext{4}{Visiting Astronomer, NOAO/CTIO, Casilla 603, La Serena, Chile}
\altaffiltext{5}{Current address: Department of Astronomy, Yale University, New Haven, CT 
06520-8101}

\begin{abstract}
We have derived Fe abundances of 16 solar-type Pleiades dwarfs by means of an equivalent width 
analysis of \ion{Fe}{1} and \ion{Fe}{2} lines in high-resolution spectra obtained with the
Hobby - Eberly Telescope and High Resolution Spectrograph.  Abundances derived from
\ion{Fe}{2} lines are larger than those derived from \ion{Fe}{1} lines (herein referred to
as over-ionization) for stars with \teff\ $< 5400$ K, and the discrepancy (\dfe\ = 
[\ion{Fe}{2}/H] - [\ion{Fe}{1}/H]) increases dramatically with decreasing \teff, reaching 
over 0.8 dex for the coolest stars of our sample.  The Pleiades joins the open clusters 
M\,34, the Hyades, IC\,2602, and IC\,2391, and the Ursa Major moving group, demonstrating 
ostensible over-ionization trends.  The Pleiades \dfe\ abundances are correlated with 
\ion{Ca}{2} infrared triplet and H$\alpha$ chromospheric emission indicators and relative 
differences therein.  Oxygen abundances of our Pleiades sample derived from the 
high-excitation \ion{O}{1} triplet have been previously shown to increase with decreasing 
\teff, and a comparison with the \dfe\ abundances suggests that the 
over-excitation (larger abundances derived from high excitation lines relative to low
excitation lines) and over-ionization effects that have been observed in cool open 
cluster and disk field main sequence (MS) dwarfs share a common origin.  Curiously, a 
correlation between the Pleiades \ion{O}{1} abundances and chromospheric emission 
indicators does not exist.  Star-to-star \ion{Fe}{1} abundances have low internal scatter 
($< 0.11$ dex), but the abundances of stars with \teff $< 5400$ K are systematically 
higher compared to the warmer stars.  The cool star [\ion{Fe}{1}/H] abundances cannot be 
connected directly to over-excitation effects, but similarities with the \dfe\ and 
\ion{O}{1} triplet trends suggest the abundances are dubious.  Using the [\ion{Fe}{1}/H] 
abundances of five stars with \teff\ $> 5400$ K, we derive a mean Pleiades cluster 
metallicity of [Fe/H] $= +0.01 \pm 0.02$.
\end{abstract}

\keywords{open clusters and associations:individual (Pleiades) --- stars:abundances ---
stars:atmospheres --- stars:late-type}

\section{INTRODUCTION}
Studies of Galactic chemical evolution are dependent on accurately derived abundances of
stars spanning all ages, populations, kinematics, masses, and metallicities.  Stars with
masses $M \leq 1 \mathrm{M}_{\odot}$ are especially important given their dominance of 
the initial mass function \citep[IMF; e.g.,][]{2002Sci...295...82K}.  Abundance 
studies utilizing high-resolution spectroscopy and local thermodynamic equilibrium (LTE) 
analyses of near-solar metallicity G and K dwarfs in open clusters and in the disk field, 
however, have revealed that the observed abundances of at least some elements derived for 
these cool main sequence (MS) dwarfs may be spurious.  In particular, studies have found 
evidence of over-ionization and over-excitation\footnotemark[6], i.e., larger abundances 
are derived from lines of singly ionized species compared to neutral species and from 
high-excitation lines of neutral species compared to low-excitation lines, respectively.  
The first indication that some abundances derived for cool MS stars are problematic may 
have come from \citet{1974ApJS...27..405O}, who found the over-ionization of Sc, Ti, Cr, 
and Fe for a sample of 10 K dwarfs ($4800 \leq$ \teff $\leq 5600$ K) in the solar 
neighborhood.  After a careful analysis of the procedures and stellar parameters used in 
the abundance derivations, the author was unable to account for the overabundances of the 
ionized species.  \citet{1998A&AS..129..237F} found similar over-ionization results for 
Sc, V, Cr, Fe, and Y in five field K dwarfs ($4510 \leq$ \teff\ $\leq 4833$ K).  The 
authors could not exclude an inaccurate temperature scale that is several hundred K too 
low as a possible source of the anomalous abundances, but in the end, they suggest that 
non-LTE (NLTE) effects are the more likely cause.

\footnotetext[6]{In this paper, we use over-ionization and over-excitation to refer to 
the observed enhanced abundances derived from spectral lines of singly ionized species or 
from high excitation lines, as opposed to other common usages referring specifically to 
the non-LTE (NLTE) effects of over-ionization (the mean intensity, $J_{\nu}$, is larger 
than the Planck function, $B_{\nu}$, in lower atomic energy levels), resonance 
scattering, and photon pumping \citep[e.g.,][]{1998A&AS..129..237F,2005ARA&A..43..481A}.}

Open clusters have been important to the identification and continued study of the 
over-excitation/ionization effects, because the presumed internal chemical homogeneity of 
the clusters provides a baseline with which anomalous abundances can be compared.  
\citet{2000ApJ...533..944K} derived O abundances from the high-excitation ($\chi = 9.15$
eV) near-IR \ion{O}{1} triplet of a K dwarf in each of the the Pleiades and NGC 2264 open 
clusters, and in both cases, the abundances were highly enhanced: [O/H] $= +0.85$ and
$+0.43$, respectively.  Such high O abundances are not expected for clusters with nominal 
metallicities of [Fe/H] $= 0.00$ \citep{1988ApJ...327..389B} and -0.15 
\citep{2000ApJ...533..944K}, respectively.  Following \citet{2000ApJ...533..944K}, 
\citet{2003AJ....125.2085S} derived the abundances of MS dwarfs in the M\,34 cluster; 
over-ionization of Fe and over-excitation of Si (the abundances of which were derived 
from lines with excitation potentials in the range $5.61 \leq \chi \leq 6.19$ eV) are 
seen in the coolest stars of the sample.  Oxygen abundances of the cool M\,34 dwarfs, as 
well as cool Pleiades dwarfs, derived from the high-excitation \ion{O}{1} triplet are 
also highly enhanced \citep{2004ApJ...602L.117S}, confirming the earlier results of 
\citet{2000ApJ...533..944K}.

Subsequent to these early open cluster studies, the over-ionization of Fe has been
confirmed in the Hyades \citep{2004ApJ...603..697Y,2006AJ....131.1057S} and Ursa Major
(UMa) moving group \citep{2005PASP..117..911K}, and of Ti in the young pre-MS clusters 
IC\,2602 and IC\,2391 \citep{2009A&A...501..553D}.  Overabundances of O derived from the 
\ion{O}{1} triplet have been reported for the UMa moving group 
\citep{2005PASP..117..911K}, the Hyades \citep{2006ApJ...636..432S}, and IC\,4665 
\citep{2007ApJ...660..712S}.  Over-excitation effects have been reported for other 
elements, as well, including S in the Pleiades \citep {2004ApJ...602L.117S}, Si, Ti, Ni,
and Cr in IC\,4665 \citep{2005ApJ...635..608S}, Ni in the Hyades 
\citep{2006AJ....131.1057S}, and Ca, Ti, and Na in IC\,2602 and IC\,2391 
\citep{2009A&A...501..553D}.  Recent abundance analyses of cool field stars have also
identified over-excitation/ionization effects 
\citep{2004A&A...420..183A,2007A&A...465..271R,2008AJ....135..618C}, confirming the 
findings of earlier work.

The over-excitation/ionization abundance anomalies are not thought to represent real 
photospheric overabundances; rather, we believe that they are a signal that our knowledge 
of cool dwarf atmospheres and/or spectral line formation therein is incomplete.  As of 
yet, the source or cause of the effects has not been identified.  Systematically 
erroneous stellar parameters, e.g., an inaccurate \teff\ scale, could lead to the 
observed abundance trends, but in general, unrealistically large parameter errors would 
have to be present \citep[e.g.,][]{2005PASP..117..911K,2006ApJ...636..432S}.  
Furthermore, parameter changes made in response to the overabundances of one element 
often increase those of another \citep[e.g.,][]{2003AJ....125.2085S}.  NLTE effects have 
been suggested as the cause \citep[e.g.,][]{1998A&AS..129..237F}, but in general, the 
over-excitation/ionization effects seen in cool dwarfs are in stark contrast to extant 
NLTE calculations.  For instance, LTE analyses of the high-excitation \ion{O}{1} triplet 
in the spectra of MS dwarfs are predicted to result in increasingly discrepant abundances 
with increasing \teff\ for stars with \teff\ $> 6000$ K, requiring negative NLTE 
corrections up to 0.4 -- 0.5 dex at 6500 K for solar metallicity dwarfs 
\citep{2003A&A...402..343T, 2009A&A...500.1221F}.  Below 6000 K, the NLTE corrections are 
predicted to be $< 0.1$ dex and essentially zero below 5500 K 
\citep{2003A&A...402..343T}.  Chromospheric emission and photospheric activity (spots, 
plages, and faculae) have also been suggested sources for the abundance anomalies 
\citep[e.g.,][]{2006ApJ...636..432S}.  These inhomogeneities could produce apparent 
over-excitation/ionization effects within a strict LTE framework.

Continuing our efforts to delineate and understand the observed over-excitation/ionization
effects in cool MS dwarfs, we have derived \ion{Fe}{1} and \ion{Fe}{2} abundances of 16 
Pleiades dwarfs, 15 of which have had O abundances derived from the high-excitation 
\ion{O}{1} triplet \citep{2004ApJ...602L.117S}.  The \ion{O}{1} abundances evince a steep
increase, reaching [O/H] $\approx 1.0$ dex near 5000 K, and star-to-star dispersion below 
5500 K.  We use the newly derived Pleiades Fe abundances to investigate if the
over-excitation and over-ionization effects observed in cool MS dwarfs are related and
indeed manifestations of the same phenomenon.  Future observational studies that could 
place stringent constraints on these effects and bring us closer to discovering the 
source of the anomalous abundances are also discussed.

\section{OBSERVATIONS AND ANALYSIS}
Echelle spectra of 17 Pleiades MS dwarfs were obtained with the 9.2-m Hobby - Eberly 
Telescope (HET) and High Resolution Spectrograph (HRS) at the McDonald Observatory in 
queue-mode on 22 separate nights between 2002 August 23 UT and 2003 February 17 UT.  
These spectra have been used previously by \citet{2004ApJ...602L.117S} and 
\citet{pleiadesJ} who analyzed the $\lambda 7775$ \ion{O}{1} high-excitation triplet and 
the $\lambda 6707$ \ion{Li}{1} line, respectively, in these Pleiades dwarfs; please 
consult these papers for detailed descriptions of the observations and spectra.  Briefly, 
the HET/HRS detector is a 4096 x 4100 side-by-side CCD mosaic of two 2048 x 4100 CCDs 
with 15 $\mu$m pixels.  The spectra are characterized by a high resolution of R $\equiv 
\; \lambda/\Delta\lambda$ = 60,000, and they have  typical signal-to-noise (S/N) ratios 
of 80 -- 100.  The spectra cover the wavelengths 5095 to 8860 {\AA}.  Data reduction 
followed the typical practice of using standard {\sf IRAF} routines to remove the bias 
pattern, subtract scattered light, flat-field, and wavelength-calibrate the spectra.  The 
stars in our sample are listed in Table 1.

Nineteen \ion{Fe}{1} and seven \ion{Fe}{2} lines spanning 5793 to 7462 {\AA} have been 
analyzed in the spectra of 16 of the seventeen Pleiads in our sample; note, however, that 
not all of the lines were measurable in all of the stars.  Equivalent widths (EWs) were 
determined by fitting each line with a Gaussian profile using the one-dimensional 
spectrum analysis software package {\sf SPECTRE} \citep{1987BAAS...19.1129F}.  The EW
measurements, along with the wavelength, excitation potential ($\chi$), and transition 
probabilities ($\log gf$) of each line, are given in Tables 2 -- 4.  Atomic data of the 
lines were obtained by email query to the Vienna Atomic Line Database (VALD) 
\citep{1995A&AS..112..525P, 1999A&AS..138..119K, 1999PhScr..T83..162}.  

One star included in our Li study \citep{pleiadesJ} but not considered here is \hii\ 152. 
This star was observed on two nights separated by approximately ten months.  As discussed 
by King et al., target misidentification on one night is a concern.  On the night where it
is not a concern, the spectra were obtained with a grating setting distinct from the 
setting used for the other stars in our sample, giving a spectral coverage of 6100 to 
9800 {\AA}.  Many of the \ion{Fe}{1} and \ion{Fe}{2} lines in our linelist fall outside 
of the spectral coverage of these data, and consequently, \hii\ 152 was not included in 
our analysis here.

The \ion{Fe}{1} and \ion{Fe}{2} abundances were derived using the LTE stellar line 
analysis package MOOG \citep {1973ApJ...184..839S} and model stellar atmospheres
with convective overshoot interpolated from the standard ATLAS9 grids of R. 
Kurucz\footnotemark[7].  In previous analyses of open cluster dwarfs 
\citep[e.g.,][]{2004ApJ...602L.117S, 2006AJ....131.1057S}, we have shown that model 
atmospheres with convective overshoot produce consistent results as those without the 
overshoot approximation (NOVER).  Presently, we have also tested for a subsample of our 
stars the more updated Kurucz models without overshoot that include the most recent 
opacity distribution functions (the ODFNEW models).  The resulting \fei\ and \feii\ 
abundances are consistent with those derived using the overshoot models; the differences 
range from 0 -- 0.05 dex, with the ODFNEW-based abundances generally lower, similar to 
what was found previously for the NOVER models.  The overshoot models used here are the 
same ones used by \citet{2004ApJ...602L.117S}, from which the adopted stellar parameters 
are also taken.  One exception is \hii\ 298; for this star, we have used the updated 
dereddened $(B-V)_0$ color of \citet{pleiadesJ} to calculate new \teff, $\log g$, and 
microturbulent velocity ($\xi$) values using the relations described in 
\citet{2004ApJ...602L.117S} and interpolate a new model from the Kurucz grids.  The 
adopted stellar parameters for our sample are provided in Table 1.

\footnotetext[7]{http://kurucz.cfa.harvard.edu/grids.html}

\section{RESULTS AND DISCUSSION}
Line-by-line abundances derived for each of our Pleiades stars are listed in Tables 2 --
4, and the stellar mean abundances can be found in Table 5.  The final mean 
abundances are given relative to solar values (Table 2) derived via an EW analysis of sky 
spectra obtained with the HET/HRS as part of our observational program.  The relative 
abundances are determined on a line-by-line basis before the mean is taken; this strict
line-by-line abundance analysis ensures that the final relative abundances are 
independent of the adopted oscillator strengths.  The adopted solar 
parameters are included in Table 1.  The mean [\ion{Fe}{1}/H] abundances and the 
differences in the mean [\ion{Fe}{2}/H] and [\ion{Fe}{1}/H] abundances 
($\Delta \mathrm{Fe} =$ [\ion{Fe}{2}/H] - [\ion{Fe}{1}/H]) are plotted in Figure 1 along 
with errorbars for three representative stars-- \hii\ 263, \hii\ 2284, and \hii\ 3179-- 
that run the \teff\ range of our sample.  The errorbars denote the total internal 
uncertainties ($\sigma_{\mathrm{Total}}$) in the derived abundances.  The  
$\sigma_{\mathrm{Total}}$ uncertainties are the quadratic sum of the abundance 
uncertainties resulting from errors in the adopted stellar parameters 
\citep{2004ApJ...602L.117S} and the uncertainty in the mean abundance (Table 5).  
Abundance sensitivities to the stellar parameters were determined by individually 
altering effective temperature ($\Delta T_{\mathrm{eff}} = \pm150\;  \mathrm{K}$), 
surface gravity ($\Delta \log g = \pm0.30\; \mathrm{dex}$), and microturbulent velocity 
($\Delta \xi = \pm0.25\;  \mathrm{km \; s}^{-1}$) and are given in Table 6.  The 
$\sigma_{\mathrm{Total}}$ uncertainties in the \ion{Fe}{1} abundances for the three 
representative stars are $\pm 0.03$, $\pm 0.05$, and $\pm 0.04$ dex for \hii\ 263, \hii\ 
2284, and \hii\ 3179, respectively.  For \ion{Fe}{2}, the $\sigma_{\mathrm{Total}}$ 
uncertainties are $\pm 0.09$, $\pm 0.10$, and $\pm 0.08$, respectively.  The errorbars 
for the [\ion{Fe}{2}/H] - [\ion{Fe}{1}/H] abundances shown in Figure 1 represent the 
quadratically combined \ion{Fe}{1} and \ion{Fe}{2} individual $\sigma_{\mathrm{Total}}$ 
uncertainties.

The star-to-star [\ion{Fe}{1}/H] abundances fall within a narrow range of 0.11 dex and 
have a standard deviation in the mean of 0.04 dex.  However, as can been seen in the left 
panel of Figure 1, the [\ion{Fe}{1}/H] abundances of stars with \teff\ $< 5400$ K are 
systematically higher than those of stars at higher \teff.  The discord is verified by 
the Spearman rank correlation coefficient ($r_s = -0.7612$) at the 99.97\% confidence 
level.  In the right panel of Figure 1, \dfe\ abundances evince a dramatic increase at 
about the same \teff, 5400 K.  The \dfe\ abundances result from large overabundances of 
\ion{Fe}{2} among the cool dwarfs (Table 5), and the \dfe\ vs. \teff\ trend for the 
Pleiades is similar to those seen in M\,34 \citep{2003AJ....125.2085S}, the Hyades 
\citep{2004ApJ...603..697Y,2006AJ....131.1057S}, and the UMa moving group 
\citep{2005PASP..117..911K}.

\subsection{Over-excitation/ionization}
The increase in \dfe\ with decreasing \teff\ presented here resembles the trend of 
increasing Pleiades O abundances derived from the high-excitation \ion{O}{1} triplet 
\citep{2004ApJ...602L.117S}.  In particular, the increase in the \ion{O}{1} triplet 
abundances also begins to become significant at approximately 5400 K.  In Figure 2 \dfe\
is plotted against the triplet abundances (\otrip), and it is seen that a strong 
correlation between these two abundance anomalies does exist.  According to the linear 
correlation coefficient ($r = 0.847$), \dfe\ and \otrip\ are correlated at a greater than 
99.9\% confidence level.  Also in Figure 2 we plot the {\it residuals} in the \dfe\ and 
\otrip\ abundances.  The residuals are differences between the observed abundances and 
\teff-dependent fitted values calculated by fitting low-order (second or third) 
polynomials to the abundance versus \teff\ relations (Figure 3); the fitted values are 
determined at each stellar \teff.  This procedure effectively removes the global mass 
dependence of the abundances so that the residuals are a measure of star-to-star 
abundance scatter at a given \teff\ \citep{2000AJ....119..859K}.  Similar to the \dfe\ 
and \otrip\ abundances, their residuals are correlated, but at a slightly lower 
confidence level, 97\% ($r = 0.589$).  The strong relationship between the \dfe\ and 
\otrip\ abundances and especially their residuals suggests the anomalous abundances share 
a common origin.

Inaccurate \teff\ scales can give rise to \teff-dependent abundance trends if the scales 
are in error in a systematic way.  \citet{2004ApJ...600..946P} have raised concern as to 
the accuracy of color-temperature relations, like the one used for our Pleiades sample, 
arguing that disagreements between observed open cluster 
color-magnitude diagrams (CMDs) and theoretical isochrones based on color-temperature 
relations likely arise from systematic errors in the latter.  \citet{2007ApJ...655..233A} 
are able to obtain near-perfect agreement between the observed CMDs of four nearby open
clusters and isochrones using empirical corrections to the color-temperature relations as 
suggested by \citet{2004ApJ...600..946P}.  Inaccurate \teff\ scales, however, do not appear 
to be at the root of the over-excitation/ionization effects observed among cool open 
cluster dwarfs.  The Pleiades and Hyades \otrip\ abundances of \citet{2004ApJ...602L.117S} 
and \citet{2006ApJ...636..432S}, respectively, were derived using color-temperature 
relations {\it and} empirically corrected isochrones, and in both cases, the steep trends 
of increasing abundances with decreasing \teff\ are present.  The \teff\ from the 
empirically corrected isochrones for the majority of dwarfs in both clusters are higher 
than those from the color-temperature relations, with the differences reaching a maximum of 
about 190 K.  Temperature corrections of this magnitude also do not alleviate the large 
\dfe\ abundances of the coolest stars in our sample.  According to the abundance 
sensitivities given in Table 6, the \teff\ of \hii\ 263 (\teff\ $= 5048$ K) would have to 
be higher by approximately 750 K in order to bring its \ion{Fe}{2} and \ion{Fe}{1} 
abundances into agreement; such errors in our adopted temperature scale are not expected 
\citep{2004ApJ...600..946P}.  Furthermore, increasing the \teff\ of \hii\ 263 by 750 K 
would result in an 0.15 dex increase in its \ion{Fe}{1} abundance and exacerbate the 
disagreement in the [\ion{Fe}{1}/H] abundances of the cool and warm dwarfs.

Whatever the cause of the anomalous \dfe\ abundances, the phenomenon may also be affecting 
the [\ion{Fe}{1}/H] abundances of the Pleiads with \teff\ $< 5400$ K.  
\citet{2003AJ....125.2085S} found that the \ion{Si}{1} abundances of the two coolest M\,34 
dwarfs ($\sim 4750$ K) they analyzed are higher by about 0.1 dex than those of the rest 
of the sample; the \ion{Si}{1} lines have excitation potentials of $\chi \sim 6$ eV.  
Similarly, Hyades \ion{Ni}{1} abundances derived from lines with excitation potentials of 
approximately 4.25 eV increase with decreasing \teff\ for dwarfs with \teff\ $\leq 5100$ 
K \citep{2006ApJ...636..432S}.  More interestingly, the \ion{Ni}{1} abundances of a 
single cool Hyades dwarf (\teff\ $= 4573$ K) derived from lines with excitation 
potentials of $\chi \approx 4.25$ eV were approximately 0.15 dex higher than the 
abundance derived from a line with a low excitation potential ($\chi = 1.83$ eV).  For a 
warmer dwarf (\teff\ $= 5978$ K), consistent abundances were obtained from all of the 
lines (please see Figure 7 of Schuler et al.).  Similar behavior is not seen here in the 
line-by-line \ion{Fe}{1} abundances of individual Pleiads (Figure 4).  However, 
identifying such excitation potential-related effects for a given cool Pleiades star is 
difficult, because the standard deviation, a measure of the dispersion in the 
line-by-line abundances, in the mean [\ion{Fe}{1}/H] abundance of each Pleiad ranges from 
0.04 to 0.13 dex and has an average of 0.07 dex.  This is of the order of the effect seen
among the cooler stars in M\,34 and Hyades dwarfs, for which the effect is expected to be
more severe.  Thus, a direct connection between the heightened mean [\ion{Fe}{1}/H] 
abundances and the over-excitation phenomenon cannot be made, but nonetheless, the fact 
that the [\ion{Fe}{1}/H] abundances increase at the same \teff\ as \dfe\ and the 
\ion{O}{1} triplet abundances is intriguing and suggests that they all are the result of 
the same effect.

\subsection{\dfe\ and Stellar Activity}
Intercluster comparisons of the cool cluster dwarf abundance anomalies can provide
valuable insight into the nature of the over-excitation/ionization effects by potentially 
linking differences in abundance trend morphologies to differences in the physical 
characteristics of the clusters, such as age and metallicity.  In Figure 5 the Pleiades 
\dfe\ values along with those of the Hyades from \citet{2006AJ....131.1057S} are plotted
versus \teff.  The \dfe\ abundances of these two clusters follow the same trend down to a 
\teff\ of about 5200 K, below which the Pleiades abundances clearly diverge.  Similar 
behavior is seen in the \ion{O}{1} abundances of dwarfs in the Hyades, Pleiades and UMa 
moving group \citep{2006ApJ...636..432S}.  The \ion{O}{1} abundances of stars in all 
three associations increase at similar rates down to a \teff\ of about 5200 K, below 
which the Pleiades trend becomes much steeper than both of those of the Hyades and UMa, 
which continue to track each other.  The UMa moving group has an age that is comparable 
to that of the Hyades \citep{2005PASP..117..911K} and a metallicity that is lower than 
both Pleiades and Hyades \citep{1990ApJ...351..467B}.  The divergence of the Pleiades 
\ion{O}{1} triplet abundances from those of both the Hyades and UMa suggests that the 
abundance trends may undergo an age-related diminution; the \dfe\ abundances of the 
Pleiades and Hyades are consistent with this conclusion.

Chromospheric emission and photospheric spots are two age-related phenomena that have been
discussed in the literature as possible sources of the observed over-excitation/ionization
effects.  \citet{2006ApJ...636..432S} demonstrated using
multicomponent model atmospheres that spotted photospheres can plausibly account for the
\ion{O}{1} triplet abundances of the cool Hyades dwarfs.  Results from efforts
investigating a possible connection between chromospheric activity and the anomalous
\ion{O}{1} triplet abundances of cool dwarfs, on the other hand, have been mixed.  No
correlation between H$\alpha$ and \ion{Ca}{2} infrared triplet emission measures and
\ion{O}{1} triplet abundances of Pleiades dwarfs nor M\,34 dwarfs exists 
\citep{2004ApJ...602L.117S}.  However, \citet{2004A&A...423..677M} found a strong
correlation between X-ray activity indicators and Pleiades triplet abundances taken from
the literature.  There is no correlation between \ion{Ca}{2} H+K emission indicators and
the Hyades \ion{O}{1} triplet abundances \citep{2006ApJ...636..432S}.  For the young
cluster IC\,4665, \citet{2007ApJ...660..712S} show that the \ion{O}{1} abundances of the
cool dwarfs are highly correlated with both H$\alpha$ and \ion{Ca}{2} infrared triplet
emission indicators.  It is important to note that Shen et al. is the only of these
studies that derived the \ion{O}{1} abundances and chromospheric emission levels from the 
same spectra.  For the others, the measurements were made or taken from different 
sources, and thus the actual chromospheric emission level may have been different when the
spectral lines were formed.

As can be seen from these various studies, it is not clear if there is a connection 
between chromospheric emission and the over-excitation of \ion{O}{1}.  Furthermore, 
chromospheric emission is often correlated with \teff, so any correlation between 
chromospheric emission and \ion{O}{1} abundances may be masking some other 
\teff-dependent effect \citep{2006ApJ...636..432S}.  Using the chromospheric emission 
data from \citet{1993AJ....106.1059S}, we have plotted the Pleiades \dfe\ versus 
\ion{Ca}{2} infrared triplet chromospheric emission indicators (\cairt \footnotemark[8]) 
in Figure 6 and find a correlation that is significant at greater than the 99.9\% 
confidence level according to the linear correlation coefficient ($r = 0.893$).  A 
similar correlation is found for the H$\alpha$ chromospheric emission (\halpha).  These 
correlations, while suggestive of a connection between chromospheric emission and the 
over-ionization of Fe, should be viewed with caution, because \halpha\ and \cairt\ are 
also correlated with \teff\ at approximately the 93\% and 98\% confidence levels.  This 
degeneracy makes it unclear if the \dfe\ abundances, like those of \ion{O}{1}, are 
affected by chromospheric emission or some other \teff-dependent effect.

\footnotetext[8]{The $\mathrm{R}^{\prime}$ index is the ratio of the flux in the line 
core (\ion{Ca}{2} infrared triplet or H$\alpha$) to the star's bolometric flux 
\citep{1984ApJ...279..763N}.}

More importantly, residual \dfe\ abundances and residual chromospheric emission 
indicators (calculated in the same manner as the abundance residuals, i.e., they are the
differences between observed and \teff-dependent fitted values; please see Figure 3) are 
correlated at the 93\% and 99.9\% confidence levels for H$\alpha$ and \ion{Ca}{2} 
infrared triplet, respectively (Figure 6).  These relationships are more indicative of a 
true correlation between chromospheric emission and the over-ionization of Fe.  We remind 
the reader, however, that the \dfe\ abundances and chromospheric emission indicators were 
not measured using the same spectra and are thus not cotemporal.

\subsection{\dfe\ Residuals and the Pleiades Li Dispersion}
\citet{pleiadesJ} have used the same HET/HRS spectra analyzed here to examine the 
long-standing problem of the large Li abundance dispersion observed among cool Pleiades 
dwarfs.  These authors find evidence that at least a portion of the dispersion is due to 
real Li depletion and suggest that the differential depletion may be a consequence of 
stellar radius modulations induced by surface magnetic activity, i.e., spots, during
pre-MS evolution.  It is also suggested that such spot-induced effects could be related 
to the over-excitation/ionization effects observed today.

\dfe\ residuals are plotted versus Li abundance residuals in Figure 7.  Li abundances are 
derived from ${\lambda}6707$ \ion{Li}{1} line strengths and are taken from 
\citet{pleiadesJ}.  The \dfe\ and Li residuals relation has a correlation coefficient of 
$r=0.46$, corresponding to a $\sim$91\% confidence level.  While only marginally 
significant at best, the mild correlation still means that a substantial fraction 
(nearly-half) of the variance in \dfe\ is related to that in Li.  The abstract picture 
painted by this result is consistent with the conjecture presented by \citet{pleiadesJ}: 
the considerable Li abundance dispersion in cool Pleiades dwarfs has a real pre-MS 
depletion component, a portion of which may be driven by the same mechanism (the 
influence of spots) but perhaps by different physics (the influence on stellar structure 
versus the influence on line formation in addition to or independent of stellar 
structure) that is also possibly responsible for the \ion{Fe}{2} versus \ion{Fe}{1} 
differences seen in these stars.  This would explain the overlap in the variance of 
Pleiades \dfe\ and Li abundances but allow for only marginally significant present-day 
correlations between these observables.  Observational tests, using the same type of 
Pleiades spectroscopic/photometric monitoring program proposed in the Conclusions section 
below, of this possibility are discussed in 
\citet{pleiadesJ}.

\subsection{Pleiades Cluster Metallicity}
The mean abundance of the Pleiades stars above 5400 K derived from \ion{Fe}{1} lines is
[Fe/H] $= +0.01 \pm 0.02$ (uncertainty in the mean) compared to [Fe/H] $= +0.07 \pm 0.01$ 
(uncertainty in the mean) for stars below 5400 K.  While a direct connection between the 
high \fei\ abundances and the over-excitation effects manifested in the \otrip\ 
abundances of cool open cluster dwarfs cannot be made here, the similarities are 
suggestive and raise doubt as to the accuracy of the cool star \fei\ abundances.  For 
this reason, and because of the anomalously high \ion{Fe}{2} abundances, we feel that the 
mean cluster metallicity is best estimated by the \fei\ abundances of the warmer stars 
only, and adopt the value of [Fe/H] $= +0.01 \pm 0.02$ for the Pleiades.

The Pleiades is one of the most well studied Galactic open clusters, and Fe abundances of 
member F, G, and K dwarfs have been derived using high-resolution spectroscopy by a 
handful of groups, which are summarized in Table 7.  \citet{1988IAUS..132..449C} found a 
mean abundance of [Fe/H] $= +0.13$ from one F and three G dwarfs, though the abundances 
are characterized by a large dispersion (0.02 -- 0.26 dex).  The spectra of three of the 
stars have S/N ratios of $\sim 40$.  In the same year, \citet{1988ApJ...327..389B} report 
a cluster mean abundance of [Fe/H] $= +0.003 \pm 0.054$ for 13 F stars that have standard 
deviations in their individual [Fe/H] abundances $\leq 0.10$ dex; the mean abundance of 
their entire 17 star sample is [Fe/H] $= -0.03$.  Subsequent to that, 
\citet{1989ApJ...336..798B} and \citet{1990ApJ...351..467B} find similar values, [Fe/H] 
$= +0.02$ and -0.02, respectively.  \citet{2000ApJ...533..944K} derived the abundances of 
two cool Pleiades K dwarfs and obtained a mean abundance of [Fe/H] $= +0.06$ from an 
analysis of \ion{Fe}{1} lines.  This value is almost identical to that of the stars with 
\teff\ $< 5400$ K in our sample, and we suspect that the abundances of the two K dwarfs 
from King et al. are similarly suspect.  However, \citet{2008AA...483..567G} also derived 
a mean abundance of [Fe/H] $= +0.06$ but for five F dwarfs.  These authors noted that 
their result is slightly larger than that of \citet{1990ApJ...351..467B} and suggested 
differences in the analyses, i.e., spectral lines used and adopted stellar parameters, as 
a possible cause.  Finally, \citet{2009PASJ...61..930F} have recently reported a cluster 
abundance of [Fe/H] $= +0.03 \pm 0.05$ based on 22 A, F, and G stars.

Excluding the results of \citet{1988IAUS..132..449C} due to poor data quality and those of
\citet{2000ApJ...533..944K} due to the uncertainty in the abundances of the two cool K
dwarfs studied, the mean Pleiades cluster abundance of the six remaining studies, 
including ours, is [Fe/H] $= +0.01$ with a standard deviation of $\sigma_{\mathrm{s.d.}} 
= 0.03$.  This is identical to the value from the five stars with \teff\ $>5400$ K in our 
sample, and by our assessment, represents the best estimate of the Pleiades cluster 
metallicity.

\section{CONCLUSIONS}
We have derived Fe abundances via an EW analysis of \ion{Fe}{1} and \ion{Fe}{2} lines in
high-resolution and moderate-S/N spectra of 16 MS dwarfs in the Pleiades open cluster. 
The \feii\ abundances increase dramatically relative to \fei\ at \teff\ below 5400 K, with
the difference reaching over 0.8 dex in the coolest stars.  This behavior is akin to what 
is seen in M\,34, the Hyades, and the UMa moving group.  Comparison of the \dfe\ abundance 
patterns in the Pleiades and Hyades, as well as the \otrip\ abundances in the Pleiades, 
Hyades, and UMa moving group, suggests that the trends may relax with age, though 
metallicity may yet prove to be a factor.  Abundances of cool dwarfs in additional open 
clusters or other stellar associations, especially those older than the Hyades, are 
needed to determine if either age or metallicity are related to these anomalous 
abundances.

The \fei\ abundances are also higher in Pleiads below 5400 K, but they show no evidence 
of an increase with decreasing \teff.  The inability to attribute the high \fei\ 
abundances of the cool stars to the over-excitation effects illustrates the difficulty of 
quantifying this phenomenon.  With lines of exceptionally high excitation potential such 
as the \ion{O}{1} triplet, the over-excitation effect is clearly seen 
\citep[e.g.,][]{2006ApJ...636..432S,2007ApJ...660..712S}, but for lines with excitation 
potentials $\lesssim 5$ eV, the effect is more difficult to pinpoint.  Our \ion{Fe}{1} 
linelist includes transitions ranging in excitation potential from 2.18 to 5.10 eV, but 
for the individual Pleiades stars, no increase in the line-by-line abundances as a 
function of excitation potential, like that seen in \ion{Ni}{1} abundances of cool 
Hyades dwarfs \citep[1.83 eV $\leq \chi \leq$ 4.42 eV;][]{2006AJ....131.1057S}, is 
evident (Figure 4).  Line-to-line sensitivities to the over-excitation/ionization effects 
have yet to be clearly delineated, and it needs to be determined if there is an 
excitation potential threshold above which the abundances derived from these lines become 
enhanced by these effects.  Similarly, it needs to be determined if there is an 
excitation potential threshold {\it below} which the opposite occurs, the abundances 
derived from the low-excitation lines are lower due to these effects.  Such behavior 
would be expected if the overabundances of high-excitation (singly ionized) lines are due 
to the overpopulation of high-excitation (singly ionized) electronic states at the 
expense of depopulating low-excitation states.  Whether or not the over-excitation 
effects impact the spectroscopic derivation of stellar parameters (\teff, $\log g$, and 
$\xi$), an approach not adopted here, also needs to be determined.  Future investigations 
of these effects will require high-quality high-resolution spectroscopy so that accurate 
line-by-line abundances can be derived, even from features of just a few m{\AA} in 
strength.

A strong correlation between the \dfe\ and \otrip\ abundances of the Pleiades dwarfs is
evident in Figure 2, suggesting that the over-excitation/ionization effects share a
common cause or origin.  Chromospheric emission and photospheric spots have been shown to
be promising culprits, but to this point, the data are inconclusive.  Whereas strong 
correlations between the Pleiades \dfe\ and chromospheric emission indicators and their 
residuals exist, they do not exist between \otrip\ and chromospheric emission.  These 
contradictory results complicate the interpretation of the observed 
over-excitation/ionization effects and will have to be addressed by future studies.  Also, 
comparing abundances and chromospheric emission indicators measured using different 
spectra may not provide an accurate test of a true correlation because of potential 
temporal changes in chromospheric activity levels.  Future investigations into the 
over-excitation/ionization effects in cool open cluster dwarfs should make every effort 
to derive chromospheric activity levels from the same spectra so that any possible 
relation between the two can be more definitively delineated.

Determining the influence of photospheric spots on abundance derivations is more arduous. 
Multicomponent model atmospheres- simulating photospheres with different areal coverages 
of hot, cool, and quiescent spots- have been shown to able to reproduce the measured EWs 
of the \ion{O}{1} triplet in a sample of Hyades stars, but, while such exercises are 
useful and demonstrate the plausibility of the photospheric spot hypothesis, the results 
are only suggestive.  Observationally, a simultaneous photometric and spectroscopic 
monitoring program could be used to identify any correlated changes in spot coverage and 
spectral line strengths.  Such observational constraints would be helpful to determine if 
spotted photospheres affect high-resolution abundance derivations.  Despite the 
challenges, the possible connection between spots, over-excitation/ionization effects, 
and pre-MS Li depletion should provide sufficient motivation for future efforts. 

A final conclusion that can be drawn from this study is that those carrying out 
spectroscopic abundance analyses of open clusters should heed caution when their samples 
include cool dwarfs, particularly those with \teff\ $< 5400$ K.  Including the abundances 
of these stars may skew cluster mean abundances.  A similar caution may be needed for 
those studying cool MS dwarfs in the disk field, as well.  Further investigations into 
the sensitivity of the over-excitation/ionization effects to excitation potential, first 
ionization potential, stellar age, and stellar metallicity are needed in order to 
identify the extent and ubiquity of these effects.

\acknowledgements
A.L.P. gratefully acknowledges her support for summer research from the National Science
Foundation through their Research Experience for Undergraduates program at Cerro Tololo
Inter-American Observatory (grant AST-0647604).  A.L.P. also thanks the C.V. 
Starr-Middlebury School in Latin America for additional support.  S.C.S. is grateful for 
support provided by the NOAO Leo Goldberg Fellowship; NOAO is operated by the Association 
of Universities for Research Astronomy, Inc., under a cooperative agreement with the 
National Science Foundation.  J.R.K. gratefully acknowledges support for this work by 
grants AST 00-86576 and AST 02-39518 to J.R.K. from the National Science Foundation and 
by a generous grant from the Charles Curry Foundation to Clemson University.

{\it Facilities:} \facility{HET (HRS)}

\newpage

\newpage
\begin{figure}
\plotone{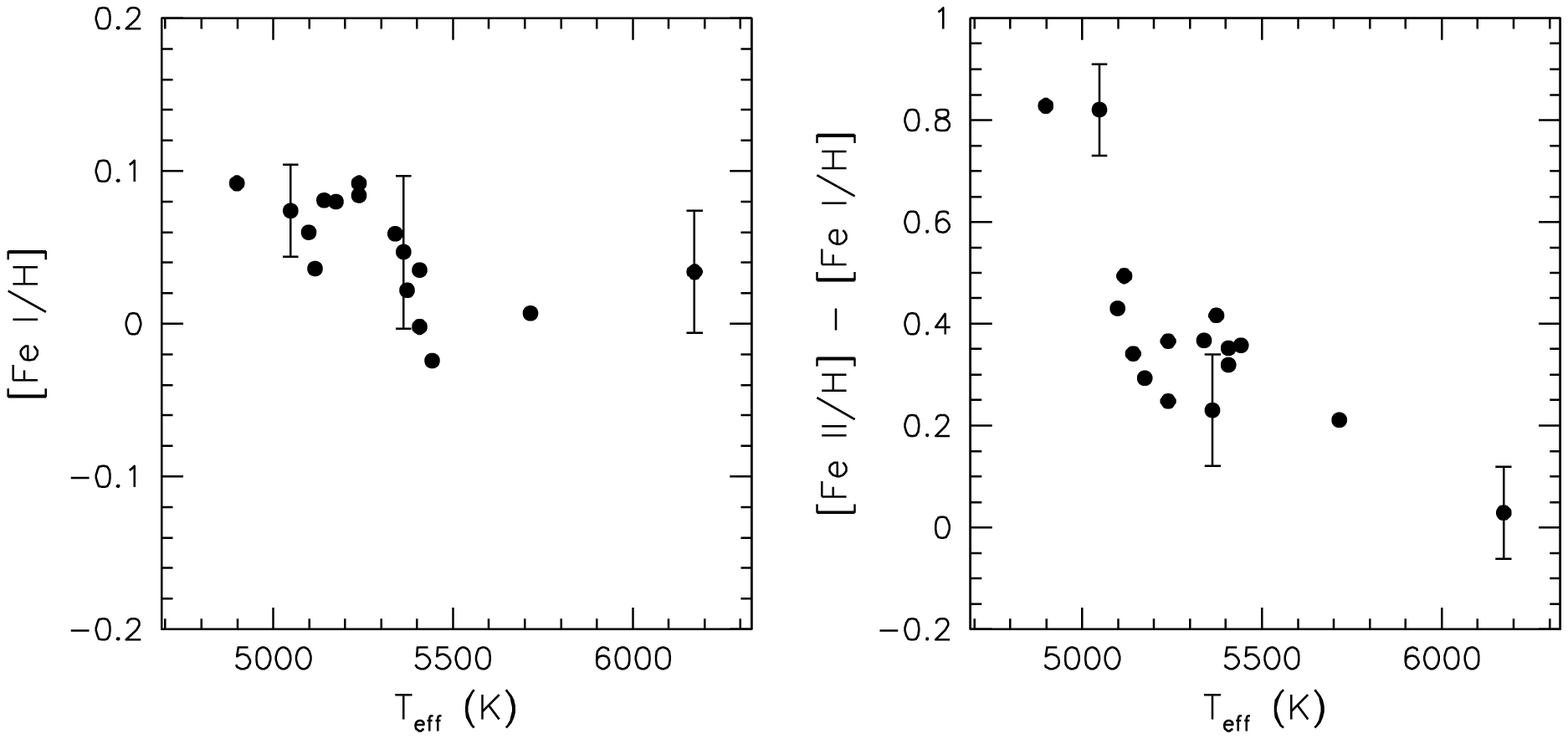}
\caption{Left: Fe abundances derived from \ion{Fe}{1} lines vs \teff.  Errorbars for three
representative stars- \hii\ 263, \hii\ 2284, and \hii\ 3179- are shown and illustrate the
total internal abundance uncertainties, as described in the text.  Right: \dfe\ vs \teff. 
The errorbars for the three representative stars depict the quadratically combined 
\ion{Fe}{1} and \ion{Fe}{2} individual total internal uncertainties.}
\end{figure}

\begin{figure}
\plotone{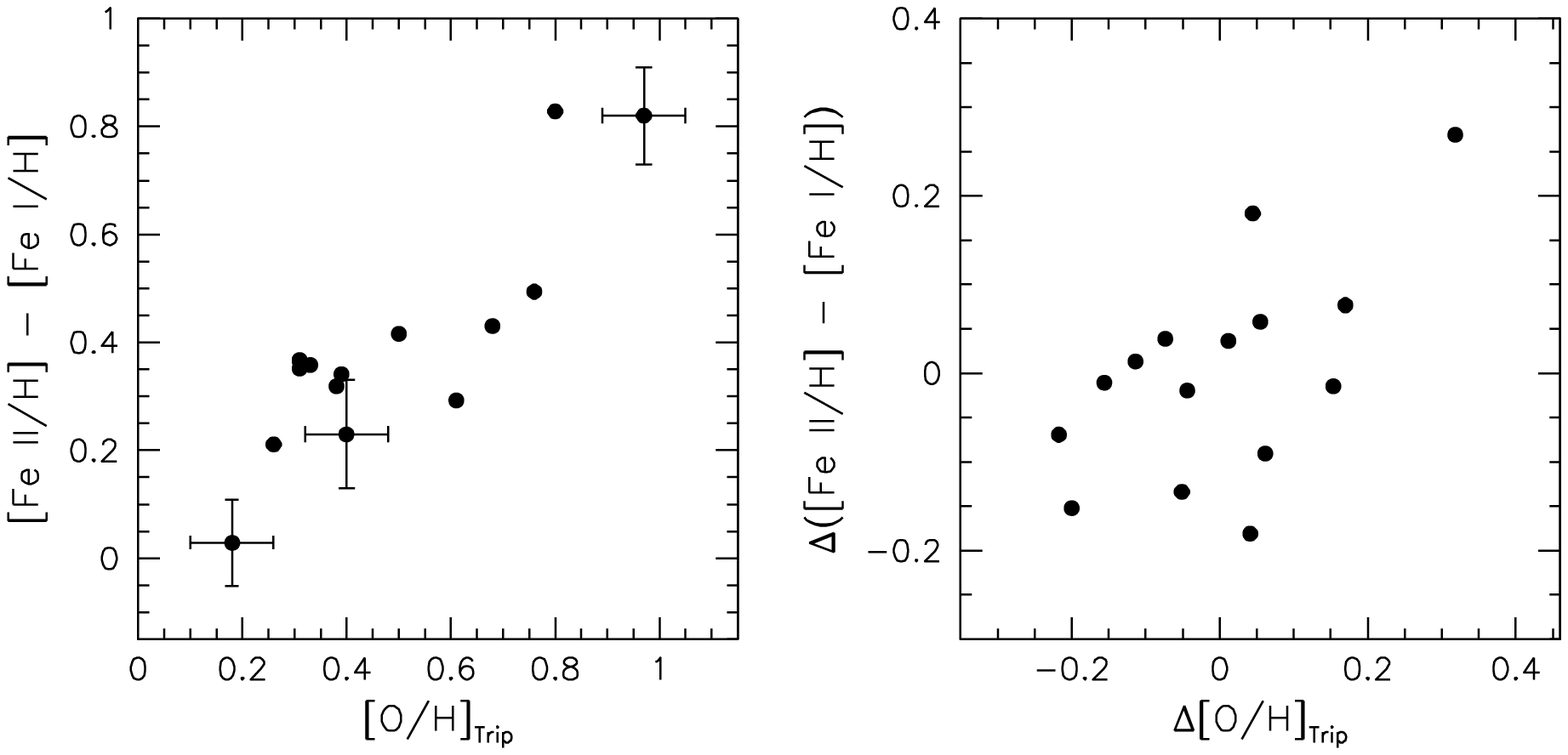}
\caption{Left: \dfe\ plotted against O abundances derived from the high-excitation 
\ion{O}{1} triplet.  The \otrip\ abundances and typical uncertainties (shown as the 
horizontal errorbars) are from \citet{2004ApJ...602L.117S}.  The vertical errorbars are 
those shown in the right panel of Figure 1.  \dfe\ is correlated with
[O/H]$_{\mathrm{Trip}}$ at greater than the 99.9\% confidence level according to the 
linear correlation coefficient ($r = 0.847$).  Right: \dfe\ residuals vs O abundance 
residuals.  The residuals are the differences in observed and \teff-dependent fitted 
values.  The residuals are correlated at a 97\% confidence level ($r = 0.589$).}
\end{figure}

\begin{figure}
\plotone{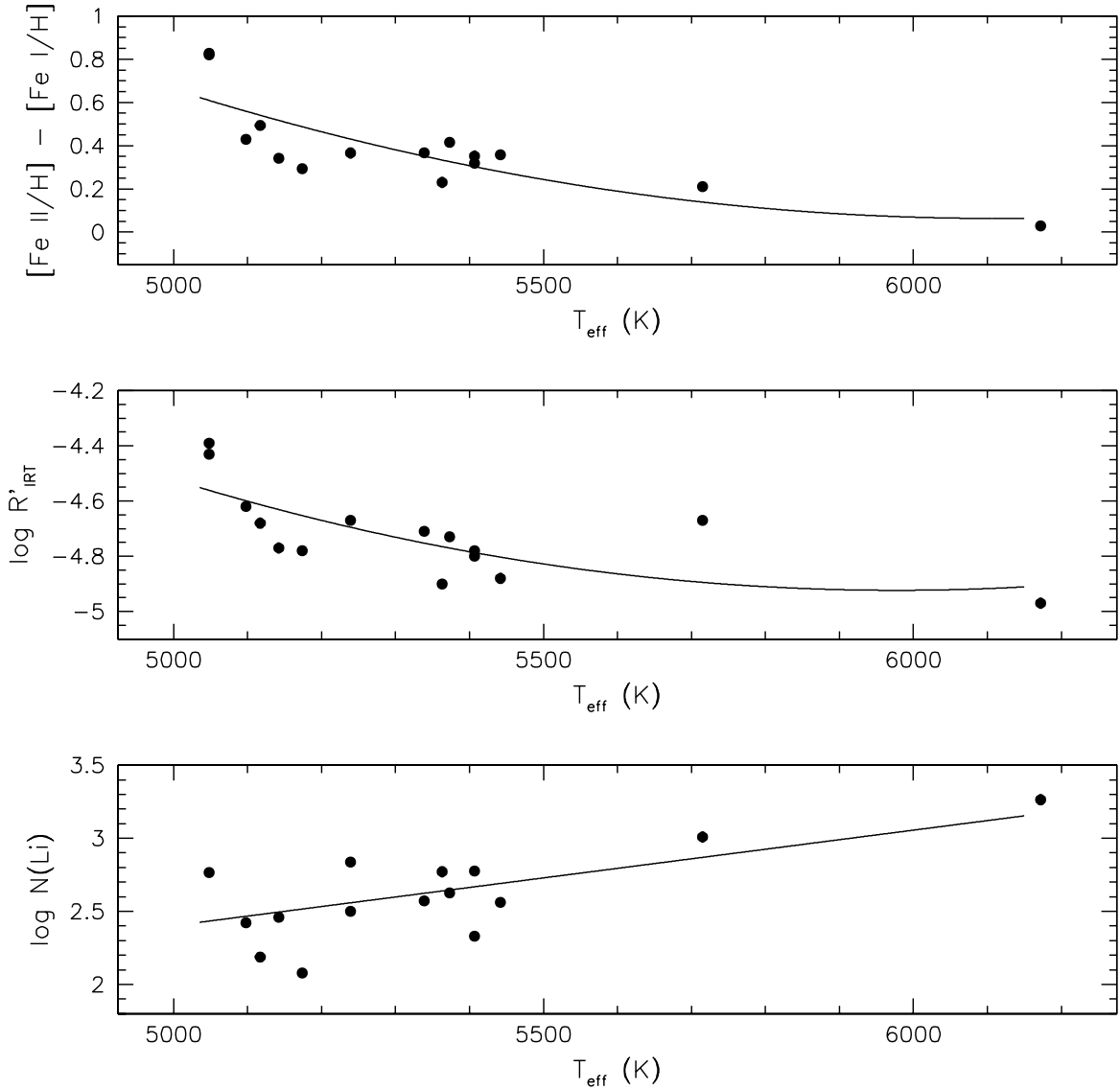}
\caption{Polynomial fits (solid lines) to \dfe, \cairt, and Li abundance versus \teff\ 
relations for the Pleiades dwarfs.  The fits are used to calculate residuals- differences 
between observed and fitted values- in the abundances and chromospheric activity 
indicators for each star.}
\end{figure}

\begin{figure}
\plotone{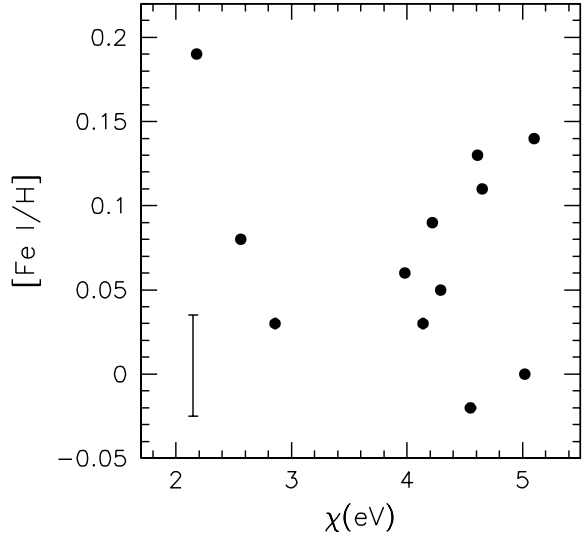}
\caption{Line-by-line Fe abundances of \hii\ 263 derived from \ion{Fe}{1} lines vs 
excitation potential.  There is no trend in the abundances as a function of excitation
potential.  The standard deviation of the abundances is $\sigma_{s.d.} = 0.06$ and is
shown as the errorbar in the lower right hand corner.}
\end{figure}

\begin{figure}
\plotone{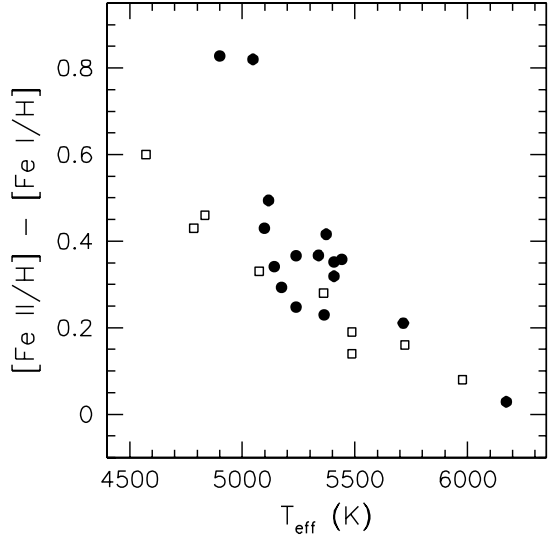}
\caption{\dfe\ for the Pleiades (closed circles) and the Hyades (open squares) as a
function of \teff.  The Hyades data are from \citet{2006AJ....131.1057S}.}
\end{figure}

\begin{figure}
\plotone{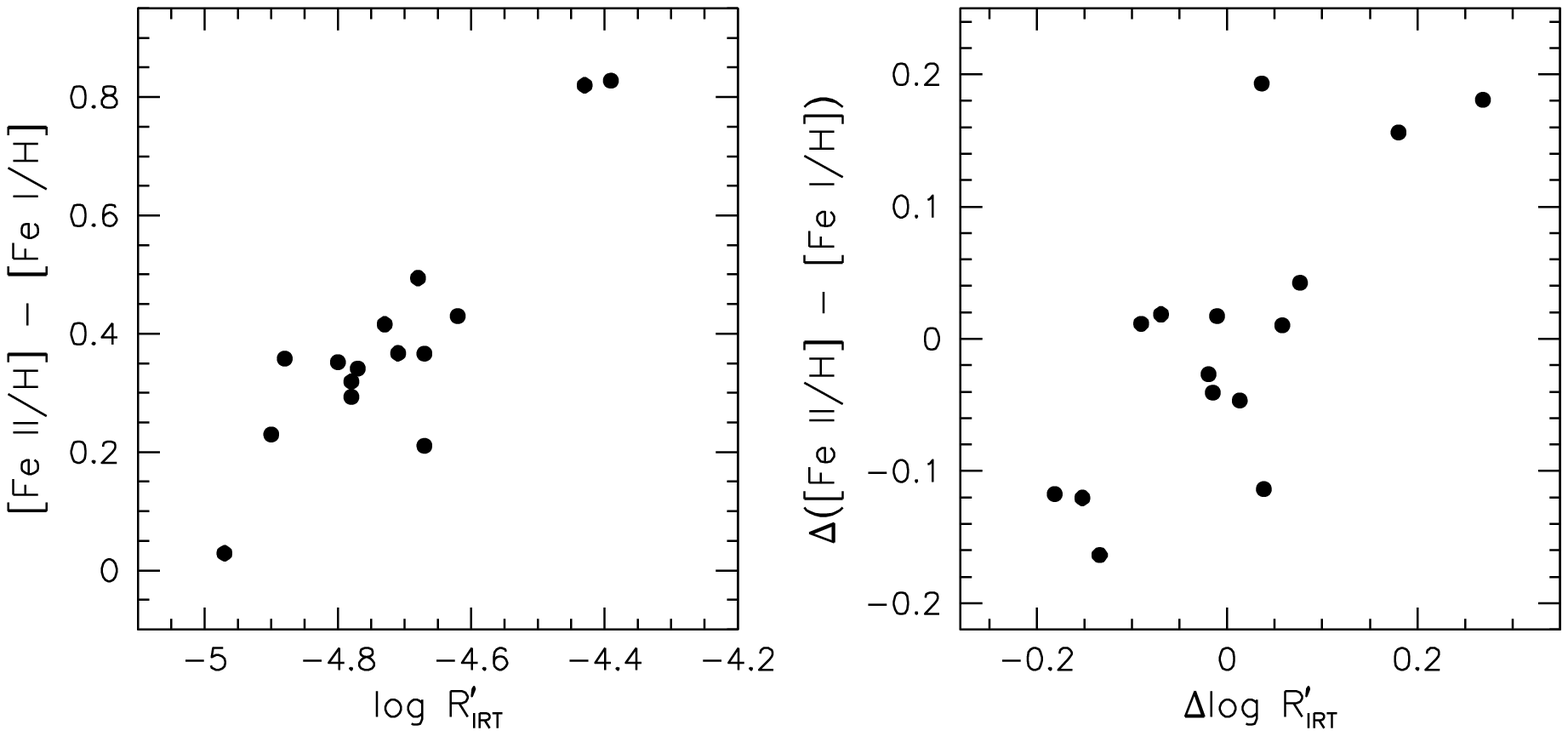}
\caption{Left: \dfe\ vs \ion{Ca}{2} infrared triplet chromospheric emission indicator.  
According to the linear correlation coefficient ($r = 0.893$), the \dfe\ abundances are
correlated with the \ion{Ca}{2} chromospheric emission with greater than 99.9\% 
confidence. The chromospheric emission data are taken from \citet{1993AJ....106.1059S}.  
Right: \dfe\ residuals vs residuals in \ion{Ca}{2} infrared triplet chromospheric 
emission indicators.  The quantities are correlated at the 99.9\% confidence level ($r = 
0.766$).}
\end{figure}

\begin{figure}
\plotone{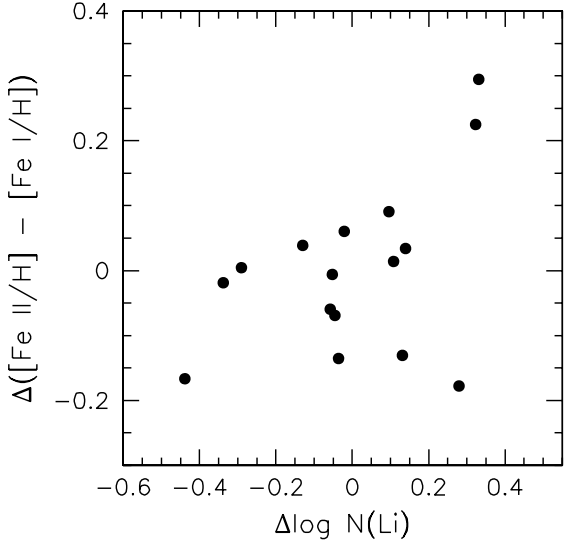}
\caption{\dfe\ residuals vs \ion{Li}{1} abundance residuals.  The \ion{Li}{1} abundances
are from \citet{pleiadesJ}.  The linear correlation coefficient ($r = 0.456$) is
marginally significant at the $\sim$91\% confidence level.}
\end{figure}

\begin{deluxetable}{lrcrrrccc}
\tablecolumns{9}
\tablewidth{0pt}
\tabletypesize{\small}
\tablecaption{Stellar Parameters}
\tablehead{
     \colhead{}&
     \colhead{}&
     \colhead{$(B-V)_o$}&
     \colhead{}&
     \colhead{$T_{\mathrm{eff}}$}&
     \colhead{}&
     \colhead{$\log g$}&
     \colhead{}&
     \colhead{$\xi$}\\
     \colhead{Star}&
     \colhead{}&
     \colhead{(mag)}&
     \colhead{}&
     \colhead{(K)}&
     \colhead{}&
     \colhead{(cgs)}&
     \colhead{}&
     \colhead{(km $\mathrm{s}^{-1}$)}
     }

\startdata
$\mathrm{HII} \; 0152$ && 0.65 && 5715 && 4.55 && 0.96\\
$\mathrm{HII} \; 0193$ && 0.75 && 5339 && 4.61 && 0.58\\
$\mathrm{HII} \; 0250$ && 0.65 && 5715 && 4.55 && 0.96\\
$\mathrm{HII} \; 0263$ && 0.84 && 5048 && 4.64 && 0.30\\
$\mathrm{HII} \; 0298$ && 0.89 && 4899 && 4.66 && 0.16\\
$\mathrm{HII} \; 0571$ && 0.75 && 5373 && 4.60 && 0.61\\
$\mathrm{HII} \; 0746$ && 0.73 && 5407 && 4.60 && 0.65\\
$\mathrm{HII} \; 0916$ && 0.82 && 5098 && 4.64 && 0.35\\
$\mathrm{HII} \; 1593$ && 0.73 && 5407 && 4.60 && 0.65\\
$\mathrm{HII} \; 2126$ && 0.81 && 5142 && 4.63 && 0.39\\
$\mathrm{HII} \; 2284$ && 0.74 && 5363 && 4.61 && 0.60\\
$\mathrm{HII} \; 2311$ && 0.78 && 5239 && 4.62 && 0.48\\
$\mathrm{HII} \; 2366$ && 0.78 && 5239 && 4.62 && 0.48\\
$\mathrm{HII} \; 2406$ && 0.72 && 5442 && 4.59 && 0.68\\
$\mathrm{HII} \; 2462$ && 0.80 && 5174 && 4.63 && 0.42\\
$\mathrm{HII} \; 2880$ && 0.82 && 5117 && 4.64 && 0.37\\
$\mathrm{HII} \; 3179$ && 0.53 && 6172 && 4.48 && 1.42\\
$\mathrm{Sun}\dotfill$&&\nodata&& 5777 && 4.44 && 1.38\\
\enddata

\end{deluxetable}

\begin{deluxetable}{lcccccccccccccccccccccccc}
\tablecolumns{25}
\tablewidth{0pt}
\tabletypesize{\footnotesize}
\rotate
\tablecaption{Equivalent Widths and Abundances}
\tablehead{
     \colhead{}&
     \colhead{}&
     \colhead{$\lambda$}&
     \colhead{}&
     \colhead{$\chi$}&
     \colhead{}&
     \colhead{}&
     \colhead{}&
     \multicolumn{2}{c}{Sun}&
     \colhead{}&
     \multicolumn{2}{c}{\hii\ 193}&
     \colhead{}&
     \multicolumn{2}{c}{\hii\ 250}&
     \colhead{}&
     \multicolumn{2}{c}{\hii\ 263}&
     \colhead{}&
     \multicolumn{2}{c}{\hii\ 298}&
     \colhead{}&
     \multicolumn{2}{c}{\hii\ 571}\\
     \cline{9-10} \cline{12-13} \cline{15-16} \cline{18-19} \cline{21-22} \cline{24-25}\\
     \colhead{Ion}&
     \colhead{}&
     \colhead{({\AA})}&
     \colhead{}&
     \colhead{(eV)}&
     \colhead{}&
     \colhead{$\log gf$}&
     \colhead{}&
     \colhead{EW}&
     \colhead{$\log N$}&
     \colhead{}&
     \colhead{EW}&
     \colhead{$\log N$}&
     \colhead{}&
     \colhead{EW}&
     \colhead{$\log N$}&
     \colhead{}&
     \colhead{EW}&
     \colhead{$\log N$}&
     \colhead{}&
     \colhead{EW}&
     \colhead{$\log N$}&
     \colhead{}&
     \colhead{EW}&
     \colhead{$\log N$}
     }

\startdata
\ion{Fe}{1}\dotfill && 5793.92 && 4.22 && -1.70 && 33.9 & 7.52 && 46.0 & 7.63 && 37.0 & 7.59 && 47.5 & 7.61 && 45.4 & 7.56 && 39.3 & 7.51\\
                    && 5856.10 && 4.29 && -1.64 && 36.6 & 7.57 && 42.1 & 7.56 && 30.8 & 7.46 && 47.8 & 7.62 && 51.1 & 7.67 && 43.0 & 7.59\\
                    && 5927.80 && 4.65 && -1.09 && 45.7 & 7.52 && 54.0 & 7.58 && 41.5 & 7.46 && 60.3 & 7.63 && 63.9 & 7.67 && 56.5 & 7.63\\
                    && 6089.57 && 5.02 && -0.94 && 41.8 & 7.64 && 47.1 & 7.63 && 35.3 & 7.52 && 49.5 & 7.64 && 56.6 & 7.74 && 43.0 & 7.57\\
                    && 6093.65 && 4.61 && -1.50 && 32.0 & 7.63 && 39.6 & 7.67 && 30.4 & 7.60 && 45.9 & 7.76 && 42.5 & 7.69 && 39.4 & 7.68\\
                    && 6096.67 && 3.98 && -1.93 && 41.2 & 7.64 && 52.8 & 7.75 && 42.4 & 7.68 && 52.9 & 7.70 && 55.1 & 7.72 && 49.2 & 7.69\\
                    && 6151.62 && 2.18 && -3.29 && 52.5 & 7.45 && 66.5 & 7.58 && 53.6 & 7.50 && 76.0 & 7.64 &&\nodata&\nodata&& 61.4 & 7.48\\
                    && 6165.36 && 4.14 && -1.47 && 48.3 & 7.46 && 58.3 & 7.56 && 47.2 & 7.47 && 58.1 & 7.49 && 57.6 & 7.47 && 53.0 & 7.46\\
                    && 6270.23 && 2.86 && -2.71 && 56.2 & 7.59 && 70.8 & 7.77 && 57.6 & 7.67 && 68.8 & 7.62 && 75.3 & 7.69 &&\nodata&\nodata\\
                    && 6627.56 && 4.55 && -1.68 && 30.9 & 7.71 && 30.6 & 7.59 && 30.0 & 7.69 && 37.0 & 7.69 && 39.1 & 7.74 && 36.6 & 7.73\\
                    && 6806.86 && 2.73 && -3.21 && 36.4 & 7.57 && 53.1 & 7.72 && 37.8 & 7.60 &&\nodata&\nodata&&\nodata&\nodata&& 47.1 & 7.61\\
                    && 6839.84 && 2.56 && -3.45 && 37.2 & 7.65 &&\nodata&\nodata&& 35.8 & 7.62 &&\nodata&\nodata&& 58.4 & 7.76 && 38.5 & 7.48\\
                    && 6842.69 && 4.64 && -1.32 && 41.8 & 7.64 &&\nodata&\nodata&& 43.5 & 7.68 &&\nodata&\nodata&&\nodata&\nodata&& 44.1 & 7.60\\
                    && 6857.25 && 4.07 && -2.15 && 23.4 & 7.54 && 35.0 & 7.66 && 23.8 & 7.54 &&\nodata&\nodata&& 37.1 & 7.66 && 28.5 & 7.52\\
                    && 6861.94 && 2.42 && -3.89 && 21.5 & 7.61 &&\nodata&\nodata&&\nodata&\nodata&&\nodata&\nodata&& 42.5 & 7.69 && 32.5 & 7.63\\
                    && 6862.50 && 4.56 && -1.57 && 29.9 & 7.58 &&\nodata&\nodata&&\nodata&\nodata&&\nodata&\nodata&& 45.9 & 7.76 && 40.2 & 7.69\\
                    && 7284.84 && 4.14 && -1.75 && 38.8 & 7.51 &&\nodata&\nodata&& 46.0 & 7.66 &&\nodata&\nodata&& 57.9 & 7.64 && 51.0 & 7.65\\
                    && 7461.53 && 2.56 && -3.58 && 29.2 & 7.59 && 37.4 & 7.53 &&\nodata&\nodata && 48.5 & 7.67 && 50.4 & 7.68 && 42.0 & 7.64\\
                    && 7547.90 && 5.10 && -1.35 && 22.6 & 7.67 && 27.5 & 7.70 &&\nodata&\nodata && 33.3 & 7.81 &&\nodata&\nodata&& 29.4 & 7.74\\
\ion{Fe}{2}\dotfill && 5264.81 && 3.23 && -3.13 && 43.8 & 7.40 && 42.1 & 7.82 && 42.9 & 7.54 && 49.9 & 8.31 &&\nodata&\nodata&& 44.0 & 7.84\\
                    && 5414.05 && 3.22 && -3.65 && 30.7 & 7.61 &&\nodata&\nodata&&\nodata&\nodata&&\nodata&\nodata&&\nodata&\nodata&&\nodata&\nodata\\
                    && 5425.25 && 3.20 && -3.39 && 39.4 & 7.53 && 39.7 & 7.99 && 40.8 & 7.71 && 41.3 & 8.32 && 39.6 & 8.45 && 40.5 & 7.97\\
                    && 6084.10 && 3.20 && -3.88 && 22.0 & 7.59 &&\nodata&\nodata&&\nodata&\nodata&&\nodata&\nodata&&\nodata&\nodata&&\nodata&\nodata\\
                    && 6247.56 && 3.89 && -2.44 && 52.6 & 7.49 && 47.4 & 7.92 && 56.8 & 7.77 && 52.5 & 8.37 && 47.2 & 8.43 && 50.5 & 7.96\\
                    && 6432.68 && 2.89 && -3.69 && 42.0 & 7.55 && 39.8 & 7.96 && 41.8 & 7.70 && 48.1 & 8.48 && 41.5 & 8.50 && 37.2 & 7.86\\
                    && 6456.39 && 3.90 && -2.19 && 63.4 & 7.45 && 55.7 & 7.86 && 69.6 & 7.79 && 67.1 & 8.41 && 53.0 & 8.32 && 63.0 & 7.98\\

\enddata

\end{deluxetable}

\begin{deluxetable}{lcccccccccccccccccccccccc}
\tablecolumns{25}
\tablewidth{0pt}
\tabletypesize{\footnotesize}
\rotate
\tablecaption{Equivalent Widths and Abundances}
\tablehead{
     \colhead{}&
     \colhead{}&
     \colhead{$\lambda$}&
     \colhead{}&
     \colhead{$\chi$}&
     \colhead{}&
     \colhead{}&
     \colhead{}&
     \multicolumn{2}{c}{\hii\ 746}&
     \colhead{}&
     \multicolumn{2}{c}{\hii\ 916}&
     \colhead{}&
     \multicolumn{2}{c}{\hii\ 1593}&
     \colhead{}&
     \multicolumn{2}{c}{\hii\ 2126}&
     \colhead{}&
     \multicolumn{2}{c}{\hii\ 2284}&
     \colhead{}&
     \multicolumn{2}{c}{\hii\ 2311}\\
     \cline{9-10} \cline{12-13} \cline{15-16} \cline{18-19} \cline{21-22} \cline{24-25}\\
     \colhead{Ion}&
     \colhead{}&
     \colhead{({\AA})}&
     \colhead{}&
     \colhead{(eV)}&
     \colhead{}&
     \colhead{$\log gf$}&
     \colhead{}&
     \colhead{EW}&
     \colhead{$\log N$}&
     \colhead{}&
     \colhead{EW}&
     \colhead{$\log N$}&
     \colhead{}&
     \colhead{EW}&
     \colhead{$\log N$}&
     \colhead{}&
     \colhead{EW}&
     \colhead{$\log N$}&
     \colhead{}&
     \colhead{EW}&
     \colhead{$\log N$}&
     \colhead{}&
     \colhead{EW}&
     \colhead{$\log N$}
     }

\startdata
\ion{Fe}{1}\dotfill && 5793.92 && 4.22 && -1.70 &&  40.4 & 7.54 && 48.0 & 7.62 && 40.5 & 7.54 && 49.5 & 7.66 && 48.2 & 7.69 && 46.5 & 7.62 \\
                    && 5856.10 && 4.29 && -1.64 &&  40.9 & 7.56 && 46.7 & 7.61 && 40.9 & 7.56 && 49.4 & 7.66 && 41.0 & 7.23 && 44.3 & 7.58 \\
                    && 5927.80 && 4.65 && -1.09 &&  53.0 & 7.57 && 58.0 & 7.60 && 51.2 & 7.54 && 60.5 & 7.64 && 53.2 & 7.56 && 55.5 & 7.58 \\
                    && 6089.57 && 5.02 && -0.94 &&  42.4 & 7.56 && 49.5 & 7.64 && 41.6 & 7.55 && 47.0 & 7.60 && 46.7 & 7.63 && 54.4 & 7.73 \\
                    && 6093.65 && 4.61 && -1.50 &&  38.5 & 7.67 && 41.0 & 7.67 && 39.0 & 7.68 && 40.7 & 7.66 && 40.1 & 7.69 &&\nodata&\nodata\\
                    && 6096.67 && 3.98 && -1.93 &&  51.0 & 7.74 && 54.0 & 7.72 && 46.8 & 7.65 && 55.2 & 7.76 && 56.1 & 7.83 && 51.3 & 7.70\\
                    && 6151.62 && 2.18 && -3.29 &&  63.9 & 7.55 && 69.5 & 7.53 && 61.7 & 7.50 && 80.0 & 7.75 && 70.9 & 7.68 && 72.9 & 7.66\\
                    && 6165.36 && 4.14 && -1.47 &&  53.5 & 7.48 && 62.3 & 7.57 && 49.9 & 7.41 && 56.6 & 7.48 && 51.6 & 7.43 && 57.0 & 7.51\\
                    && 6270.23 && 2.86 && -2.71 &&  63.5 & 7.64 && 67.0 & 7.60 && 64.2 & 7.66 && 73.1 & 7.73 && 71.2 & 7.54 && 69.5 & 7.70\\
                    && 6627.56 && 4.55 && -1.68 &&  35.1 & 7.70 && 41.1 & 7.78 && 35.1 & 7.70 && 38.6 & 7.73 && 42.2 & 7.84 && 39.6 & 7.76\\
                    && 6806.86 && 2.73 && -3.21 &&  46.3 & 7.60 && 55.1 & 7.68 && 44.6 & 7.57 && 58.7 & 7.77 && 49.8 & 7.66 && 53.7 & 7.70\\
                    && 6839.84 && 2.56 && -3.45 &&  42.5 & 7.58 && 52.0 & 7.67 && 39.1 & 7.51 && 46.9 & 7.57 && 43.8 & 7.59 && 49.6 & 7.66\\
                    && 6842.69 && 4.64 && -1.32 &&  51.0 & 7.73 && 48.5 & 7.64 && 48.6 & 7.69 &&\nodata&\nodata&& 49.9 & 7.70 && 48.1 & 7.65 \\
                    && 6857.25 && 4.07 && -2.15 &&  33.2 & 7.64 && 33.0 & 7.57 && 31.1 & 7.59 && 35.3 & 7.63 && 30.2 & 7.56 && 37.6 & 7.69 \\
                    && 6861.94 && 2.42 && -3.89 &&  26.6 & 7.51 && 33.4 & 7.53 && 26.3 & 7.50 &&\nodata&\nodata&& 32.0 & 7.61 &&\nodata&\nodata \\
                    && 6862.50 && 4.56 && -1.57 &&  36.4 & 7.62 && 42.5 & 7.70 && 35.0 & 7.59 &&\nodata&\nodata&& 43.0 & 7.74 && 40.3 & 7.66  \\
                    && 7284.84 && 4.14 && -1.75 &&  50.8 & 7.65 && 55.6 & 7.69 && 48.8 & 7.61 &&\nodata&\nodata&& 48.8 & 7.60 &&\nodata&\nodata \\
                    && 7461.53 && 2.56 && -3.58 &&  45.0 & 7.72 && 48.4 & 7.68 && 36.6 & 7.54 && 42.2 & 7.55 && 45.9 & 7.72 && 47.6 & 7.71 \\
                    && 7547.90 && 5.10 && -1.35 && \nodata&\nodata&&\nodata&\nodata&&\nodata&\nodata&& 31.7 & 7.78 && 27.0 & 7.69 && 31.7 & 7.78 \\
\ion{Fe}{2}\dotfill && 5264.81 && 3.23 && -3.13 &&  37.7 & 7.64 && 36.6 & 7.91 && 41.9 & 7.76 && 32.7 & 7.75 && 34.7 & 7.60 && 37.7 & 7.79 \\
                    && 5414.05 && 3.22 && -3.65 && \nodata&\nodata&&\nodata&\nodata&&\nodata&\nodata&&\nodata&\nodata&&\nodata&\nodata && 27.0 & 7.98 \\
                    && 5425.25 && 3.20 && -3.39 &&  37.3 & 7.86 &&\nodata&\nodata&& 38.0 & 7.88 && 32.1 & 7.96 && 39.6 & 7.96 && 37.3 & 8.01 \\
                    && 6084.10 && 3.20 && -3.88 && \nodata&\nodata&&\nodata&\nodata&& 21.3 & 7.88 &&\nodata&\nodata&& 14.5 & 7.66 &&\nodata&\nodata \\
                    && 6247.56 && 3.89 && -2.44 &&  51.0 & 7.93 && 38.4 & 7.96 && 48.5 & 7.87 && 40.7 & 7.97 && 44.0 & 7.81 && 43.8 & 7.93 \\
                    && 6432.68 && 2.89 && -3.69 &&  36.8 & 7.82 && 30.5 & 7.95 && 38.6 & 7.86 && 29.9 & 7.88 && 35.2 & 7.82 &&\nodata&\nodata \\
                    && 6456.39 && 3.90 && -2.19 &&  62.4 & 7.94 && 51.2 & 8.03 && 58.7 & 7.86 && 50.6 & 7.97 && 54.5 & 7.82 && 57.8 & 8.02 \\

\enddata

\end{deluxetable}

\begin{deluxetable}{lccccccccccccccccccccc}
\tablecolumns{22}
\tablewidth{0pt}
\tabletypesize{\footnotesize}
\rotate
\tablecaption{Equivalent Widths and Abundances}
\tablehead{
     \colhead{}&
     \colhead{}&
     \colhead{$\lambda$}&
     \colhead{}&
     \colhead{$\chi$}&
     \colhead{}&
     \colhead{}&
     \colhead{}&
     \multicolumn{2}{c}{\hii\ 2366}&
     \colhead{}&
     \multicolumn{2}{c}{\hii\ 2406}&
     \colhead{}&
     \multicolumn{2}{c}{\hii\ 2462}&
     \colhead{}&
     \multicolumn{2}{c}{\hii\ 2880}&
     \colhead{}&
     \multicolumn{2}{c}{\hii\ 3179}\\
     \cline{9-10} \cline{12-13} \cline{15-16} \cline{18-19} \cline{21-22}\\
     \colhead{Ion}&
     \colhead{}&
     \colhead{({\AA})}&
     \colhead{}&
     \colhead{(eV)}&
     \colhead{}&
     \colhead{$\log gf$}&
     \colhead{}&
     \colhead{EW}&
     \colhead{$\log N$}&
     \colhead{}&
     \colhead{EW}&
     \colhead{$\log N$}&
     \colhead{}&
     \colhead{EW}&
     \colhead{$\log N$}&
     \colhead{}&
     \colhead{EW}&
     \colhead{$\log N$}&
     \colhead{}&
     \colhead{EW}&
     \colhead{$\log N$}
     }

\startdata
\ion{Fe}{1}\dotfill && 5793.92 && 4.22 && -1.70 && 47.8  & 7.65  && 43.1  & 7.61  && 48.0  & 7.64  && 43.5  & 7.54   && 26.3    & 7.58\\
                    && 5856.10 && 4.29 && -1.64 && 46.8  & 7.63  && 38.9  & 7.52  && 46.6  & 7.62  && 46.8  & 7.61   && 24.8    & 7.55\\
                    && 5927.80 && 4.65 && -1.09 && 57.3  & 7.61  && 50.5  & 7.54  && 58.3  & 7.61  && 55.2  & 7.55   && 34.2    & 7.52\\
                    && 6089.57 && 5.02 && -0.94 && 51.4  & 7.69  && 40.2  & 7.53  && 49.5  & 7.65  && 48.6  & 7.63   && 30.4    & 7.60\\
                    && 6093.65 && 4.61 && -1.50 && 47.2  & 7.80  && 32.6  & 7.55  && 41.2  & 7.68  && 43.5  & 7.72   && 24.1    & 7.66\\
                    && 6096.67 && 3.98 && -1.93 && 49.8  & 7.67  && 45.5  & 7.64  && 56.1  & 7.78  && 55.1  & 7.75   && 29.5    & 7.65\\
                    && 6151.62 && 2.18 && -3.29 && 72.5  & 7.66  && 55.4  & 7.38  && 70.5  & 7.59  && 73.2  & 7.61   && 36.8    & 7.51\\
                    && 6165.36 && 4.14 && -1.47 && 57.3  & 7.51  && 51.3  & 7.45  && 56.5  & 7.48  && 55.9  & 7.46   && 37.3    & 7.50\\
                    && 6270.23 && 2.86 && -2.71 && 71.3  & 7.74  && 61.2  & 7.61  && 72.9  & 7.74  && 65.9  & 7.59   && 44.8    & 7.70\\
                    && 6627.56 && 4.55 && -1.68 && 34.8  & 7.66  && 36.0  & 7.73  && 39.6  & 7.75  && 30.2  & 7.55   && 21.5    & 7.70\\
                    && 6806.86 && 2.73 && -3.21 && 56.8  & 7.76  && 43.5  & 7.56  && 56.3  & 7.73  && 57.1  & 7.73   && 23.7    & 7.62\\
                    && 6839.84 && 2.56 && -3.45 && 51.6  & 7.71  && 42.5  & 7.60  && 47.9  & 7.60  && 49.3  & 7.62   &&\nodata  &\nodata\\
                    && 6842.69 && 4.64 && -1.32 && 48.5  & 7.65  && 44.5  & 7.62  && 49.5  & 7.67  && 52.8  & 7.72   && 32.5    & 7.67\\
                    && 6857.25 && 4.07 && -2.15 && 38.2  & 7.71  &&\nodata&\nodata&& 33.3  & 7.59  && 35.5  & 7.63   && 17.2    & 7.60\\
                    && 6861.94 && 2.42 && -3.89 && 41.1  & 7.76  &&\nodata&\nodata&& 36.9  & 7.64  && 36.6  & 7.61   && 11.0    & 7.60\\
                    && 6862.50 && 4.56 && -1.57 &&\nodata&\nodata&&\nodata&\nodata&& 44.5  & 7.74  && 38.4  & 7.62   && 25.3    & 7.68\\
                    && 7284.84 && 4.14 && -1.75 &&\nodata&\nodata&&\nodata&\nodata&& 57.5  & 7.73  &&\nodata&\nodata &&\nodata  &\nodata\\
                    && 7461.53 && 2.56 && -3.58 && 42.4  & 7.59  && 34.2  & 7.51  && 44.8  & 7.62  && 42.0  & 7.54   && 18.9    & 7.67\\
                    && 7547.90 && 5.10 && -1.35 &&\nodata&\nodata&& 24.3  & 7.64  &&\nodata&\nodata&& 30.1  & 7.74   &&\nodata  &\nodata\\
\ion{Fe}{2}\dotfill && 5264.81 && 3.23 && -3.13 && 34.0  & 7.69  && 37.8  & 7.61  && 36.9  & 7.84  && 38.9  & 7.95   && 51.4    & 7.48\\
                    && 5414.05 && 3.22 && -3.65 && 23.5  & 7.86  &&\nodata&\nodata&& 21.6  & 7.86  &&\nodata&\nodata && 30.3    & 7.53\\
                    && 5425.25 && 3.20 && -3.39 && 33.8  & 7.91  && 36.1  & 7.79  && 30.8  & 7.88  && 34.8  & 8.07   && 48.1    & 7.63\\
                    && 6084.10 && 3.20 && -3.88 &&\nodata&\nodata&&\nodata&\nodata&&\nodata&\nodata&& 39.4  & 7.97   && 26.5    & 7.64\\
                    && 6247.56 && 3.89 && -2.44 && 41.0  & 7.86  && 51.3  & 7.90  && 38.0  & 7.86  && 34.5  & 8.05   && 66.6    & 7.64\\
                    && 6432.68 && 2.89 && -3.69 &&\nodata&\nodata&& 42.4  & 7.93  && 31.5  & 7.90  && 52.0  & 8.03   && 48.9    & 7.61\\
                    && 6456.39 && 3.90 && -2.19 && 51.2  & 7.87  && 60.8  & 7.86  && 50.8  & 7.93  &&\nodata&\nodata && 74.1    & 7.53\\

\enddata

\end{deluxetable}

\begin{deluxetable}{rrrrrrrrrrr}
\tablecolumns{1}
\tablewidth{0pt}
\tablecaption{Mean Iron Aubndnaces}
\tablehead{
     \colhead{Star}&
     \colhead{}&
     \colhead{[\ion{Fe}{1}/H]}&
     \colhead{}&
     \colhead{$\sigma_{\mu}$\tablenotemark{a}}&
     \colhead{}&
     \colhead{[\ion{Fe}{2}/H]}&
     \colhead{}&
     \colhead{$\sigma_{\mu}$}&
     \colhead{}&
     \colhead{$\Delta$Fe\tablenotemark{b}}
     }

\startdata
\hii\  193 &&  0.06 && 0.02 && 0.43 && 0.01 && +0.37\\
\hii\  250 &&  0.01 && 0.02 && 0.22 && 0.05 && +0.21\\
\hii\  263 &&  0.07 && 0.02 && 0.89 && 0.04 && +0.82\\
\hii\  298 &&  0.09 && 0.01 && 0.92 && 0.02 && +0.83\\
\hii\  571 &&  0.02 && 0.02 && 0.44 && 0.04 && +0.42\\
\hii\  746 &&  0.04 && 0.02 && 0.35 && 0.06 && +0.31\\
\hii\  916 &&  0.06 && 0.01 && 0.49 && 0.05 && +0.43\\
\hii\ 1593 &&  0.00 && 0.01 && 0.35 && 0.02 && +0.35\\
\hii\ 2126 &&  0.08 && 0.03 && 0.42 && 0.04 && +0.34\\
\hii\ 2284 &&  0.05 && 0.03 && 0.28 && 0.06 && +0.23\\
\hii\ 2311 &&  0.08 && 0.01 && 0.45 && 0.04 && +0.37\\
\hii\ 2366 &&  0.09 && 0.02 && 0.34 && 0.04 && +0.25\\
\hii\ 2406 && -0.02 && 0.01 && 0.33 && 0.05 && +0.35\\
\hii\ 2462 &&  0.08 && 0.02 && 0.37 && 0.04 && +0.29\\
\hii\ 2880 &&  0.04 && 0.02 && 0.53 && 0.02 && +0.49\\
\hii\ 3179 &&  0.03 && 0.01 && 0.06 && 0.03 && +0.03\\
\enddata

\tablenotetext{a}{$\sigma_{\mu} = \sigma_{s.d.} / \sqrt{N-1}$, where $N$ is the number of
lines measured.}
\tablenotetext{b}{$\Delta$ Fe = [\ion{Fe}{2}/H] - [\ion{Fe}{1}/H].}

\end{deluxetable}

\begin{deluxetable}{llllrrr}
\tablecolumns{7}
\tablewidth{0pt}
\tablecaption{Abundance Sensitivities}
\tablehead{
     \colhead{Star}&
     \colhead{}&
     \colhead{Parameter}&
     \colhead{}&
     \colhead{\ion{Fe}{1}}&
     \colhead{}&
     \colhead{\ion{Fe}{2}}
     }

\startdata
\hii\ 263 $(T_{\mathrm{eff}} = 5048 \; \mathrm{K})$ && $\Delta T_{\mathrm{eff}}(\pm150\;  \mathrm{K})$   && $ \pm 0.03       $ && $\mp 0.16        $\\
                                                    && $\Delta \log g (\pm0.30\;  \mathrm{dex})$         && $ \pm 0.01       $ && $^{+0.11}_{-0.14}$\\
                                                    && $\Delta \xi  (\pm0.25\;  \mathrm{km \; s}^{-1})$  && $ \mp 0.02       $ && $^{-0.03}_{+0.01}$\\
\hii\ 2284$(T_{\mathrm{eff}} = 5363 \; \mathrm{K})$ && $\Delta T_{\mathrm{eff}} = \pm150\;  \mathrm{K}$  && $^{+0.07}_{-0.05}$ && $ \mp 0.11       $\\
                                                    && $\Delta \log g = \pm0.30\;  \mathrm{dex}$         && $ \pm 0.00       $ && $ \pm 0.14       $\\
                                                    && $\Delta \xi = \pm0.25\;  \mathrm{km \; s}^{-1}$   && $ \mp 0.03       $ && $^{-0.05}_{+0.03}$\\
\hii\ 3179$(T_{\mathrm{eff}} = 6172 \; \mathrm{K})$ && $\Delta T_{\mathrm{eff}} = \pm150\;  \mathrm{K}$  && $ \pm 0.10       $ && $ \mp 0.03       $\\
                                                    && $\Delta \log g = \pm0.30\;  \mathrm{dex}$         && $ \mp 0.01       $ && $ \pm 0.12       $\\
                                                    && $\Delta \xi = \pm0.25\;  \mathrm{km \; s}^{-1}$   && $ \mp 0.02       $ && $ \mp 0.06       $\\

\enddata

\end{deluxetable}

\begin{deluxetable}{llrll}
\tablecolumns{5}
\tablewidth{0pt}
\tablecaption{Census of Pleiades Cluster Metallicities}
\tablehead{
     \colhead{[Fe/H]}&
     \colhead{}&
     \colhead{$\sigma_{\mu}$}&
     \colhead{}&
     \colhead{Reference}
     }

\startdata
$+0.13$\tablenotemark{a} && 0.06 && \citet{1988IAUS..132..449C}\\
$-0.03$ && 0.02 && \citet{1988ApJ...327..389B}\\
$+0.02$ && 0.03 && \citet{1989ApJ...336..798B}\\
$-0.02$ && 0.02 && \citet{1990ApJ...351..467B}\\
$+0.06$\tablenotemark{a} && 0.03 && \citet{2000ApJ...533..944K}\\
$+0.06$ && 0.01 && \citet{2008AA...483..567G}\\
$+0.03$ && 0.05 && \citet{2009PASJ...61..930F}\\
$+0.01$ && 0.02 && This work\\
        &&      &&          \\
$+0.01$ && 0.01 && Mean\\

\enddata

\tablenotetext{a}{Not included in the Mean abundance, as described in the text.}

\end{deluxetable}

\end{document}